\documentclass[prl,twocolumn,superscriptaddress,nofootinbib,aps,10pt,preprintnumbers,longbibliography]{revtex4-2}
\usepackage[colorlinks=true,linkcolor=black,citecolor=blue,urlcolor=blue, pdfborder={0 0 0}]{hyperref}
 \usepackage{amssymb,amsmath,amsthm,cancel,graphicx}
\usepackage{mathtools}
\usepackage[utf8]{inputenc}
\usepackage{color}
\usepackage[table,xcdraw,dvipsnames]{xcolor}
\usepackage{multirow}
\usepackage[normalem]{ulem}    % per tagliare \sout{}
\usepackage{bm} % boldphase en mathmode
\newcommand{\Sec}[1]{ \medskip \noindent {\sl \bfseries #1}}
\definecolor{ultramarine}{RGB}{0,32,96}

\hypersetup{
    pdftoolbar=true,        
    pdfmenubar=true,        
    pdffitwindow=false,     
    pdfstartview={FitH},    
    pdftitle={Atoms as electron accelerators for 
measuring the  e+e- to hadrons cross section},
    pdfauthor={F. Arias Aragon, L. Darme, G. Grilli di Cortona, E. Nardi},
    pdfkeywords={hadronic vacuum polarization} {atomic electron velocities},
    pdfnewwindow=true,
    colorlinks=true,
    linkcolor=blue,
    citecolor=blue,
    filecolor=blue,
    urlcolor=blue,
    linktocpage=true
}

\def\eq#1{{Eq.~(\ref{#1})}}

\def\beq{\begin{equation}}  
\def\eeq{\end{equation}}
\def\beqa{\begin{eqnarray}}  
\def\eeqa{\end{eqnarray}}

\newcommand{\ba} {\begin{equation}\begin{aligned}}
\newcommand{\ea} {\end{aligned}\end{equation}}

\newcommand{\bg} {\begin{equation}\begin{gathered}}
\newcommand{\eg} {\end{gathered}\end{equation}}

\newcommand{\cM}{\mathcal{M}}
\newcommand{\cE}{\mathcal{E}}

\def\({\left(}
\def\){\right)}
\def\[{\left[}
\def\]{\right]}

\def\eg{\hbox{\it e.g.}{}}

\def\amIW{a^\text{HVP}_{\text{W}}}
\def\amhvp{a_{\mu}^{\rm HVP}}
\def\shad{\sigma_{\mathrm{had}}}
\begin{document}

\title{
Atoms as electron accelerators for 
measuring the  $e^+e^- \to\,$hadrons cross section
}

\author{Fernando~Arias-Arag\'on}
\email{fernando.ariasaragon@lnf.infn.it}
\affiliation{Istituto Nazionale di Fisica Nucleare, Laboratori Nazionali di Frascati, Frascati, 00044, Italy}

\author{Luc Darmé}
\email{l.darme@ip2i.in2p3.fr}
\affiliation{Universit\'e Claude Bernard Lyon 1, CNRS/IN2P3, Institut de Physique des 2 Infinis de Lyon, UMR 5822, F-69622, Villeurbanne, France}

\author{Giovanni Grilli di Cortona}
\email{giovanni.grilli@lngs.infn.it}
\affiliation{Istituto Nazionale di Fisica Nucleare, Laboratori Nazionali del Gran Sasso, Assergi, 67100, L'Aquila (AQ), Italy}
 
\author{Enrico Nardi}
\email{enrico.nardi@lnf.infn.it}
\affiliation{Istituto Nazionale di Fisica Nucleare, Laboratori Nazionali di Frascati, Frascati, 00044, Italy}
\affiliation{Laboratory of High Energy and Computational Physic, HEPC-NICPB, R\"avala 10, 10143, Tallin, Estonia}
\date{\today}

\begin{abstract}
The hadronic vacuum polarization contribution 
to $(g-2)_\mu$ can be determined  via dispersive methods from 
$e^+e^-\to\;$hadrons data. 
We propose a novel approach to measure  the hadronic cross section $\shad(s)$  
as an  alternative to  the initial-state radiation and energy scan techniques, 
which relies on positron annihilation off atomic electrons of a high  $Z$ target  
($^{238}$U, $Z=92$). We show that by 
leveraging the relativistic electron velocities 
 of the inner atomic shells,
a high-intensity $12\,$GeV positron beam, such as the one foreseen at JLab, 
can allow to measure $\shad(s)$  with high statistical accuracy  
from the two-pion 
threshold up to above $\sqrt{s} \sim 1\,$GeV.
\end{abstract}

\maketitle 

%%%%%%%%%%%%%%%%
%%%%%%%%%%%%%%%%

\parskip 2pt

\Sec{Introduction.}  
Accurate predictions for the muon anomalous magnetic moment $a_\mu$~\cite{Jegerlehner:2009ry, Aoyama:2020ynm}, when compared with precise experimental measurements, provide a powerful test of the Standard Model (SM). This is because all three SM sectors - QED, Weak, and QCD - contribute to determine its value.
Regrettably, while the  precise experimental determinations 
from BNL~\cite{Bennett:2006fi} and 
FNAL~\cite{Abi:2021gix,Muong-2:2023cdq} are in excellent 
agreement,  the theoretical situation remains unsatisfactory, as different evaluations yield discordant results. 

The hadronic vacuum polarization (HVP),
whose contribution to the muon anomalous magnetic moment is commonly denoted as $\amhvp$, involves nonperturbative QCD effects 
and is, by far, the most complex and least controlled input in  the theoretical calculations.
Evaluations of the HVP are carried out 
relying on two different strategies. 
From first principles, by means of QCD lattice techniques, 
and via dispersive methods - a data driven approach that uses as input the hadronic cross section $\shad$ measured in $e^+ e^-$ annihilation. 
So far, the most precise lattice result has been 
obtained by the BMW collaboration~\cite{Borsanyi:2020mff,Boccaletti:2024guq}, and no other lattice 
determinations of the full $\amhvp$ with comparable precision are yet available. 
Nevertheless, the partial contribution $\amIW$, that corresponds  
to the so-called  intermediate Euclidean time-distance window, 
in which lattice-related systematic and statistical uncertainties are under good control, has been evaluated by several other collaborations~\cite{Ce:2022kxy,Alexandrou:2022amy,Lehner:2020crt,
Aubin:2022hgm,Aubin:2019usy,Wang:2022lkq,Bazavov:2023has,Blum:2023qou}. 
The results are in good agreement among them and with
BMW, giving strong support to the reliability of lattice evaluations.

However, the BMW  result  is in tension with the 
recommended  value of $\amhvp$ derived from
$e^+e^- \to$ hadrons cross section data~\cite{Aoyama:2020ynm} measured at $e^+e^-$ circular colliders.
The tension  is exacerbated in the case of $\amIW$, 
for which the dispersive method yields a value~\cite{Colangelo:2022vok} that  is several standard deviations below the average of  lattice results.\footnote{The tension 
between the data driven result~\cite{Colangelo:2022vok} and  individual results of  different lattice collaborations is  around $4\sigma$~\cite{Borsanyi:2020mff,Ce:2022kxy,Alexandrou:2022amy,Blum:2023qou}.  
Ref.~\cite{Alexandrou:2022amy}  quotes $4.5\sigma$ 
for the combined BMW~\cite{Borsanyi:2020mff},  CLS/Mainz~\cite{Ce:2022kxy}, and  
ETMC~\cite{Alexandrou:2022amy} estimates neglecting correlations. 
Ref.~\cite{WittingMoriond2023} quotes a $3.8\sigma$ tension 
for the  combined BMW,  CLS/Mainz,   ETMC and RBC/UKQCD~\cite{Blum:2023qou} 
assuming 100\% correlation.}
The situation is further complicated by the presence of 
significant disagreements between different experimental 
determinations of $\shad$.
The two most precise determinations based on the initial-state radiation (ISR) technique by KLOE~\cite{KLOE-2:2017fda} and 
BaBar~\cite{BaBar:2009wpw, BaBar:2012bdw}, exhibit a  a long-standing $\sim\!3\sigma$ 
discrepancy.\footnote{It has been speculated recently that higher-order QED effects might play 
a role in this disagreement~\cite{BaBar:2023xiy}.} 
This disagreement has been recently overshadowed by the new CMD-3 result~\cite{CMD-3:2023alj} obtained by using the energy scan technique, which is well above (and hardly consistent with) the KLOE and BaBar determinations. For example, in the energy range $\sqrt{s} \in [0.6,\,0.88]\,$GeV 
the CMD-3 contribution to $\amhvp$ is more than $5\,\sigma$ above  
the estimate based on KLOE data. 
These discrepancies underscore the urgent need for new, accurate determinations of $\shad$. The impact of such measurements would be significantly amplified if a novel method, distinct from the ISR and scanning techniques, could be devised to determine the energy dependence of the cross section.\footnote{The MUon on Electron (MUonE) elastic scattering experiment~\cite{Abbiendi:2016xup}, which aims to determine $\amhvp$ using data from elastic muon scattering off atomic electrons in the spacelike region, is a notable example of  an alternative strategy.}

The aim of this work is to propose a new technique to measure $\shad$ which could be implemented leveraging high energy and high luminosity positron beams that could 
become readily available in the near future.
The strategy is inspired by the recent realization  
that in positron annihilation on fixed targets,  
the momentum distribution of atomic electrons 
allows to scan over a large range of 
 centre-of-mass (c.m.) energies, even when keeping  
  the  beam energy fixed~\cite{Arias-Aragon:2024qji}.\footnote{The importance 
  of accounting for atomic electron velocities 
  in positron annihilation on fixed target experiments was 
  first highlighted in Ref.~\cite{Nardi:2018cxi}.}
  In this Letter we  consider a target of natural (or depleted) uranium ($^{238}$U),  
that has the largest nuclear charge  ($Z=92$) among all natural elements.  
The relativistic  velocities of the electrons in the inner atomic shells
allow to effectively probe $\shad(s)$ at c.m. energies up to $\sqrt{s}\sim 1\,$GeV
even with beam energies $E_b \sim 12\,$GeV, that are well below the 
threshold   for $2\pi$ ($E_b = 77\,$GeV) and $2\mu$  ($E_b = 44\,$GeV) 
production for positron annihilating off electrons-at-rest.

The  letter is organised as follows:  
we first outline the steps required to account for 
the atomic electron momentum distribution $n({k})$ in evaluating the cross 
section for $2 \to 2$ scatterings, concentrating   on 
the process $e^+e^- \to \mu^+ \mu^-$. 
We next describe the methods we have used to 
evaluate $n(k)$, with particular attention to the high momentum tail 
of the distribution. Finally, we  focus on high precision  measurements of $\shad(s)$
achievable via positron annihilation on electrons of a fixed $^{238}$U target. 
To illustrate our strategy, we explore the potential reach  of the 
12\,GeV positron beam whose development is 
under study  at the Continuous Electron Beam Accelerator Facility
(CEBAF) at Jefferson Laboratory 
(JLab)~\cite{Afanasev:2019xmr,Accardi:2020swt,Arrington:2021alx},
and of a $O(100)\,$GeV high-quality positron beam 
that can be available at the SPS H4 beamline~\cite{Gatignon:2730780,poker:2024jan} in the CERN North Experimental Area~\cite{Banerjee:2774716}.
We argue that in the range $\sqrt{s} \leq 1\,$GeV  a  
statistical precision on $\shad(s)$ better than the typical benchmarks of circular 
$e^+ e^-$ colliders can be obtained  at the CEBAF facility, while for the CERN 
H4 beam line an increase of the positron beam luminosity 
by at least  three orders of magnitude 
would be required to provide a competitive measurement. 
It is important to stress that the novel  
technique of exploiting positron annihilation on fixed target 
for $\shad(s)$ measurements also has a high degree
of complementarity with   $e^+ e^-$ colliders measurements, as 
it provides the largest statistics in the  $\sqrt{s}$ region
close to threshold, where collider data are generally  affected 
by large statistical fluctuations.

\Sec{Di-muon cross section.}
Let us consider the luminosity independent measurement of $\shad$ that 
can be obtained from the experimentally measured $R$-ratio 
$R(s)~=~N_{\rm had}(s)/N_{\mu\mu}(s)$ 
(where  $N_{\rm had}$  and $N_{\mu\mu}$ denote respectively 
the number of hadronic  and di-muon events) multiplied 
by the $e^+e^- \to \mu^+\mu^-$ theoretical cross-section:\footnote{This method has been used for example by 
the BaBar collaboration~\cite{BaBar:2009wpw, BaBar:2012bdw} and by KLOE in Ref.~\cite{KLOE:2012anl}. It relies on the assumption that $\sigma_{\mu\mu}^{\rm th}$ is fully determined by SM processes (see~\cite{Darme:2021huc,Darme:2022yal} for a counterexample).}
\beq
\shad(s) =  R(s) % \frac{N_{\rm had}(s)}{N_{\mu\mu}(s)}
\;\sigma_{\mu\mu}^{\rm th}(s)\,.
\label{eq:shad}
\eeq
The specific process we are interested in is 
the annihilation of a positron with given energy $E_B$ and momentum 
$\bm{p}_B =(0,0,p_B)$ with an atomic electron 
in a certain  orbital $n,l$ 
with momentum-space wavefunction $\phi_{nl}(\bm{k}_A)$, 
yielding a di-muon final state $\mu^-(\bm{p}_1)+  \mu^+(\bm{p}_2)$. 
The differential cross section,  
denoted as $d\sigma$ for brevity, 
can be written 
in terms of the spin-averaged matrix element $|\cM|^2$  for 
 $e^+e^-\xrightarrow{}\mu^+\mu^-$ scattering 
between free  particle states as 
(see Refs.~\cite{Essig:2011nj,Essig:2015cda,Catena:2019gfa} for details): 
\ba
d\sigma =&\frac{d^3p_1 d^3p_2d^3k_A }{(2\pi)^{5}16E_1E_2}  \frac{\left|\phi_{nl}(\bm{k}_A)\right|^2\left|\cM\right|^2}{|E_Bk_A^z-E_{k_A} p_B|} \\
\times&\delta^3(\bm{k}_A+\bm{p}_B-\bm{p}_f)\delta(E_A+E_B-E_f),
\ea
where 
$\bm{k}_A$, $\bm{p}_B$, $\bm{p}_f=\bm{p}_1 + \bm{p}_2$ denote three-momenta,   $E_f = E_1+E_2$ and, neglecting binding energies in the energy conservation 
condition,  $E_A=m_e$ (see Ref.~\cite{Plestid:2024xzh,Plestid:2024jqm} for 
a dedicated study of atomic binding energy effects). 
Finally, in the denominator in the first line, $E_{k_A}=\sqrt{k_A^2+m_e^2}$
arises from proper normalization of the one particle free electron states. 
We can readily integrate over the  
three-momentum of  one muon  (e.g. $\bm{p}_1$) leaving a single delta function  $\delta(E_A+E_B-(\omega_1+E_2)) $, where  $\omega_1$ is expressed in terms of the
remaining three-momenta as: 
\beq
\omega_1 = \sqrt{ p_1^2 + m_\mu^2},\qquad  
p_1^i=k_A^i- p_2^i + p_B\, \delta_{iz} \,,
\eeq
with $i=x,y,z$. Assuming an isotropic distribution for  the electron momentum,
we can use spherical coordinates $d^3 k_A = k_A^2 d k_A d c_{\theta_A}  d \varphi_A $
and $d^3 p_2 = p_2^2 d p_2\,d c_{\theta_2} d \varphi_2$, where 
$ c_{\theta_A} = \cos{\theta_A}$ 
and $c_{\theta_2} = \cos{\theta_2}$.
We can  now rewrite 
\ba
\omega_1 =& \sqrt{ a - b \, c_{\varphi_2-\varphi_A}},\\
a =& p_2^2+m_\mu^2+k_A^2+p_B^2+2k_Ap_Bc_{\theta_A}\\
-&2p_2c_{\theta_2}(p_B+k_Ac_{\theta_A}), \\
b =&  2k_Ap_2s_{\theta_2}s_{\theta_A}.
\ea
Let us define $\cE = E_A+E_B-E_2$.
The integration over $\varphi_2$ and $\varphi_A$   can be performed analytically, after rewriting 
\beq
\frac{\delta(\omega_1 - \cE)}{\omega_1}  =
2\; \frac{\delta\left(\varphi_2-\varphi_2^+\right)+\delta\left(\varphi_2-\varphi_2^-\right)}{b \sqrt{1 - d^2} } \,, 
\eeq
where $\varphi_2^\pm = \varphi_A \pm \arccos{d} $,   
$ d={(a-\cE^2)}/{b}$. This yields
\beq
\!\!\!
\int_0^{2\pi}\!\!\!\!\! d\varphi_A\!\!
\int_0^{2\pi} \!\!\!\!\!d\varphi_2  \frac{\delta(\omega_1 - \cE)}{\omega_1}  = 
\frac{4 \left(2\pi-\arccos{d}\right)}{b \sqrt{1 - d^2}  }\,\Pi\left(\frac{d}{2}\right)
\eeq
with $\Pi\left(x\right) =1 $ for $|x| \leq \frac{1}{2}$ and $0$ otherwise.
Changing  variable from  $c_{\theta_A}$ to $ s =2m_e^2+2(E_B E_A -p_Bk_A c_{\theta_A})$, we 
finally obtain: 
\ba
% \nonumber
\label{eq:d2sigma}
\frac{d^2 \sigma}{d s \,dc_{\theta_2}}  = 
& \int_{m_\mu}^\infty \!\!  dE_2  
\int_0^\infty \!\! dk_A \; \frac{\left|\cM\right|^2} {32 \pi^2}
\; \frac{\left|\phi_{nl}(k_A)\right|^2}{16 \pi^3} 
\\
\times  & \frac{  (2\pi-\arccos{d})\, \Pi\left(\frac{d}{2}\right)}
{p_B |E_Bk_A c_{\theta_A}-E_{k_A}p_B|\, s_{\theta_2}s_{\theta_A}\, \sqrt{1 - d^2}} \,,
\ea
where  $c_{\theta_A} = (2 m^2_e+2E_AE_B - s - k_A^2)/(2p_Bk_A)$ and $s_{\theta_A} = (1-c_{\theta_A}^2)^{1/2}$.  
The $\Pi$ function, which restricts 
the integration to values for which $|d|<1$ 
(that in turn implies $k_A$ larger than a certain
$k_A^\mathrm{min}$)  enforces implicitly energy conservation.

The squared free matrix element reads 
\ba
|\cM|^2=&\frac{32\pi^2\alpha^2}{(m_\mu^2+ P_1\cdot P_2)^2}\biggl\{m_e^2 \bigl[2m_\mu^2+(P_1\cdot P_2)\bigr] \\
+& m_\mu^2(K_A\cdot P_B) + (K_A\cdot P_1)(P_B\cdot P_2) \\ 
+&(K_A\cdot P_2)(P_B\cdot P_1) \biggr\},
\ea
where capital letters denote free particle four-momenta (in particular 
$K_A=(E_{k_A},\bm{k}_A$)) subject to the three-momentum conservation condition~\cite{Essig:2011nj}.

\Sec{Electron momentum distribution.}
The differential cross section 
for dimuon production in positron annihilation off $^{238}$U atomic electrons 
is obtained by summing \eq{eq:d2sigma} over all electron orbitals, 
that is by replacing $|\phi_{n\ell}(k_A)|^2$
with the properly normalised (isotropic) electron momentum distribution:
\begin{equation}
    n(k_A)= \sum_{n\ell} |\phi_{n\ell}(k_A)|^2, \quad
    % \int   \frac{d^3k_A}{(2\pi)^3}  n(\bm{k}_A) = 
 \int   \frac{k^2_A  n(k_A) dk_A}{2\pi^2} = Z\,,
\end{equation}
with $Z=92$. For any given material, the electron momentum distributions   
is related to its isotropic Compton Profile (CP) $J(k)$ via   
\begin{equation}
    n(k) = -\frac{(2\pi)^2}{k}\frac{dJ(k)}{dk}.
\end{equation}
CPs are directly measurable in photon scattering experiments, and provide 
an important confirmation of \emph{ab initio} theoretical calculations.
For our estimates, we adopt the theoretical CP for $^{238}$U given in  Ref.~\cite{BIGGS1975201}, where orbital and total-atom CPs are computed for heavy elements using relativistic Dirac-Hartree-Fock wavefunctions, up to $k=100\,\mathrm{a.u.}\,\simeq 370\,$keV. 
For  momenta above $370\,$keV and  up to $k_A\simeq 11\,$MeV we complement the results of~\cite{BIGGS1975201}  by numerically estimating the contribution of the core orbitals  up to the $3d$ shell, 
 using the code DBSR-HF~\cite{ZATSARINNY2016287}. For even larger momenta we use a simple 
approximation (valid for $k_A\gg m_e\,\alpha\,Z$) which only includes the contribution of the electrons in the $1s$ shell
\begin{align}
        n^{1s} (k_A)   = \frac{32\,\pi\,(1+1.174\,Z^2\,\alpha^2)}{Z \alpha  \, m_e^3  \, x_A^2 (1+x_A^2)^{1+\sqrt{1-Z^2\alpha^2}}}\, ,
\end{align}
where  $x_A=\frac{k_A}{m_e \alpha Z}$ is the reduced momentum and 
$Z=92$ is the (unscreened)  nuclear charge for $1s$ electrons 
localized close to the nucleus.
%\bigskip
Finally, we cut off the  momentum distribution at $k_A \lesssim m_\mu$
which ensures that all the muons produced are propagating
in the forward direction and  can thus be detected. In practice,  
because of the strong suppression of the tail of the electron momentum distribution, up to $\sqrt{s} \sim 1\,$GeV$^2$ our results are insensitive to the precise choice 
of the cutoff, as long as it remains above $\sim m_\mu/2$.\footnote{Note that $k_A \sim m_\mu/2$ corresponds to a length of about  half the  $^{238}$U nuclear radius $R_U \simeq 7.4\,$fm~\cite{Blatt:1952ije}. While we expect  
that our estimate of muon production remains reliable 
also at nuclear-size distances, hadron production might receive 
corrections from nuclear effects.
}

\Sec{Statistical procedure.}
The HVP contribution to $a_\mu$ is related to the hadronic cross section via
the dispersion integral
\begin{equation}
    \amhvp = \frac{1}{4 \pi^3}\int_{4m_\pi^2}^\infty ds\, \shad(s) \,K(s),
    %\label{eq:amuHVP}
\end{equation}
with $K(s) = \int_0^1 dz \frac{z^2(1-z)}{z^2+s(1-z)/m_\mu^2}$. 
The integral is estimated from experimental data by replacing it 
with a finite sum over bins of width $\Delta s_i$:
\begin{equation}
    \amhvp \simeq \frac{1}{4\pi^3} \sum_i^{n_{\mathrm{bin}}} \Delta s_i \, \shad(s_i) \,K(s_i) \equiv \sum_i^{n_{\mathrm{bin}}} a_\mu^i, % (s_i),
\end{equation}
where $\sigma_{\pi\pi}(s_i)$ is the cross section %
at $s_i$. Given that  $\shad$  is expressed in terms of the $R$-ratio  (see~\eq{eq:shad}) 
the statistical uncertainty corresponds to the sum in quadrature of the 
uncertainties on $N^i_{\rm had}$ and  $N^i_{\mu\mu}$ in each bin:

\begin{equation}
    \delta \amhvp 
     = 
    \sqrt{\sum_i^{n_{\mathrm{bin}}} (a_\mu^i)^2 \left[ \left( \frac{\delta N_{\pi\pi}^i}{N_{\pi\pi}^i}\right)^2 + \left( \frac{\delta N_{\mu\mu}^i}{N_{\mu\mu}^i}\right)^2 \right]}
\end{equation}
where
\ba
\frac{\delta N_{\mu\mu}(s_i)}{N_{\mu\mu}(s_i)} =& \frac{1}{\sqrt{N_{\mu\mu}(s_i)}}\,, \\
    \frac{\delta N_{\pi\pi}(s_i)}{N_{\pi\pi}(s_i)} =& \frac{1}{\sqrt{R(s_i) N_{\mu\mu}}}.
\ea

\begin{figure}[t!]
    \centering
        \includegraphics[width=0.49 \textwidth]{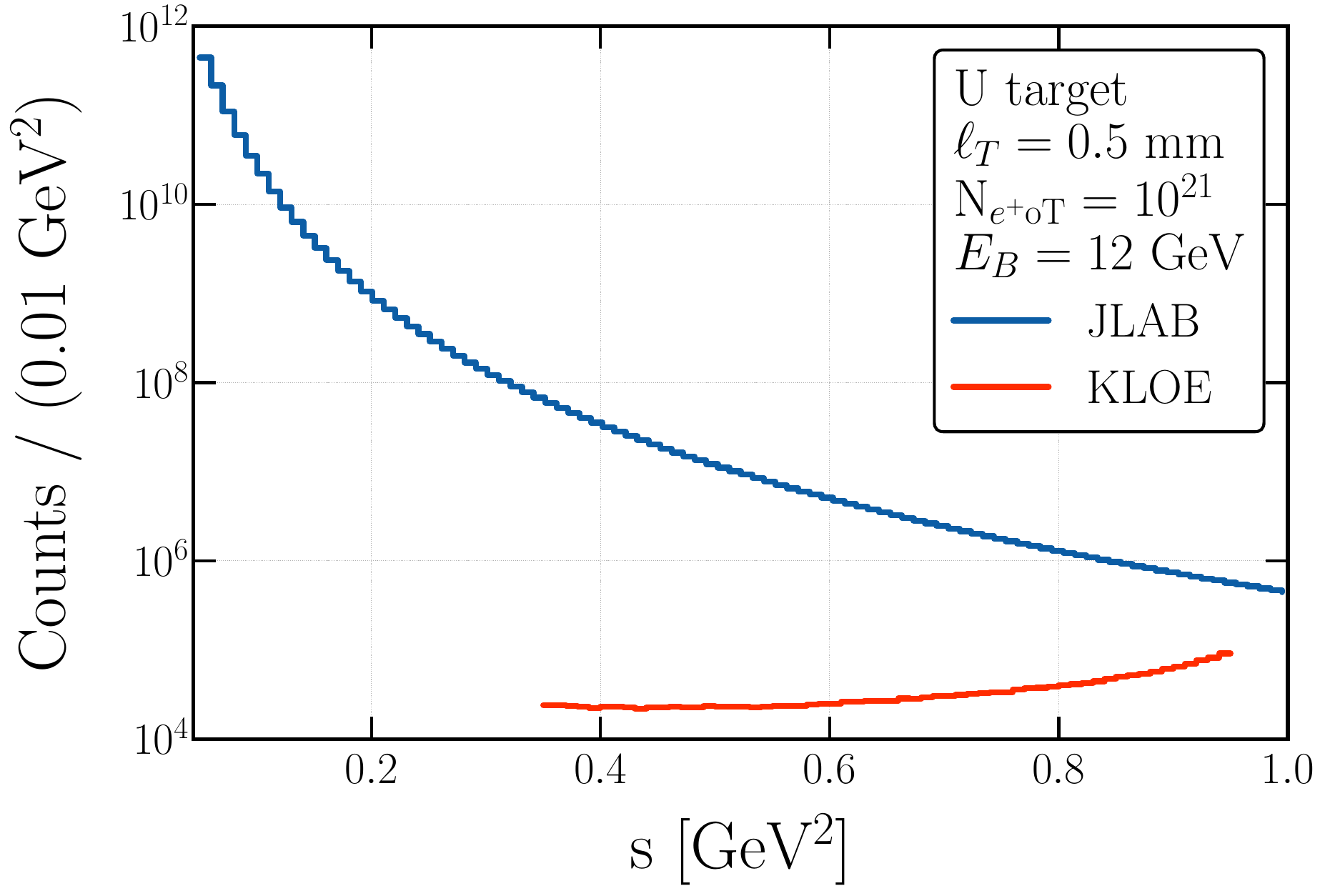}
    \caption{Number of $\mu^+\mu^-$ events produced at  JLAB with  a beam energy of $E_B=12$ GeV, compared with the number of $\mu^+\mu^-(\gamma)$ events detected by KLOE, as reported 
    in Ref.~\cite{KLOE:2012anl}.  
    }
    \label{fig:JLAB}
\end{figure}

\Sec{Positron beams and projections.}
The CEBAF injector at JLab is anticipated to be capable of producing 
 $1-5\,\mu$A unpolarized positron beams, and acceleration to energies up to $11-12\,$GeV.
The  higher figures quoted are demanding but still realistic~\cite{e.voutier}. 
In our study, we have taken $E_B = 12\,$GeV with a negligible energy spread, 
$10^{21}$ positrons on target ($e^+$oT) 
corresponding to one year of data taking with a $5\,\mu$A positron current, and a thin 
uranium target of thickness $\ell_T=500\,\mu$m. Given that the U radiation length is 
$X_0^U \simeq 3.2$\,mm,  this allows us to neglect $e^+$ energy losses inside the target. 
It should be remarked, however, that if the positron current, or the duration of the data taking run, are decreased by a factor of a few, this could be effectively compensated by 
increasing the target thickness by similar factors.
Our results are shown in Figure~\ref{fig:JLAB}. The blue line corresponds to the 
number of  di-muon events, binned in intervals of equal size 
$\Delta s_i= 0.01\,$GeV$^2$, that can be produced at JLab. 
The red line gives for comparison the number of  $\mu^+\mu^-(\gamma)$ events detected 
by   KLOE, as reported in Ref.~\cite{KLOE:2012anl}. It is apparent that 
even at c.m. energies of the order of 1\,GeV, where the suppression 
from the electron momentum  distribution becomes particularly strong, 
JLab has still the potentiality of collecting a statistics larger than those of circular $e^+e^-$ colliders. As a result, the  statistical procedure that we have outlined,  applied to the simulated JLAB dataset in the range $4m_\pi^2< s < 1\;\mathrm{GeV}^2$, 
yields a statistical uncertainty below $0.01\%$.  This clearly implies that 
the measurement will be dominated by systematic uncertainties.

\begin{figure}[t!]
    \centering
        \includegraphics[width=0.49 \textwidth]{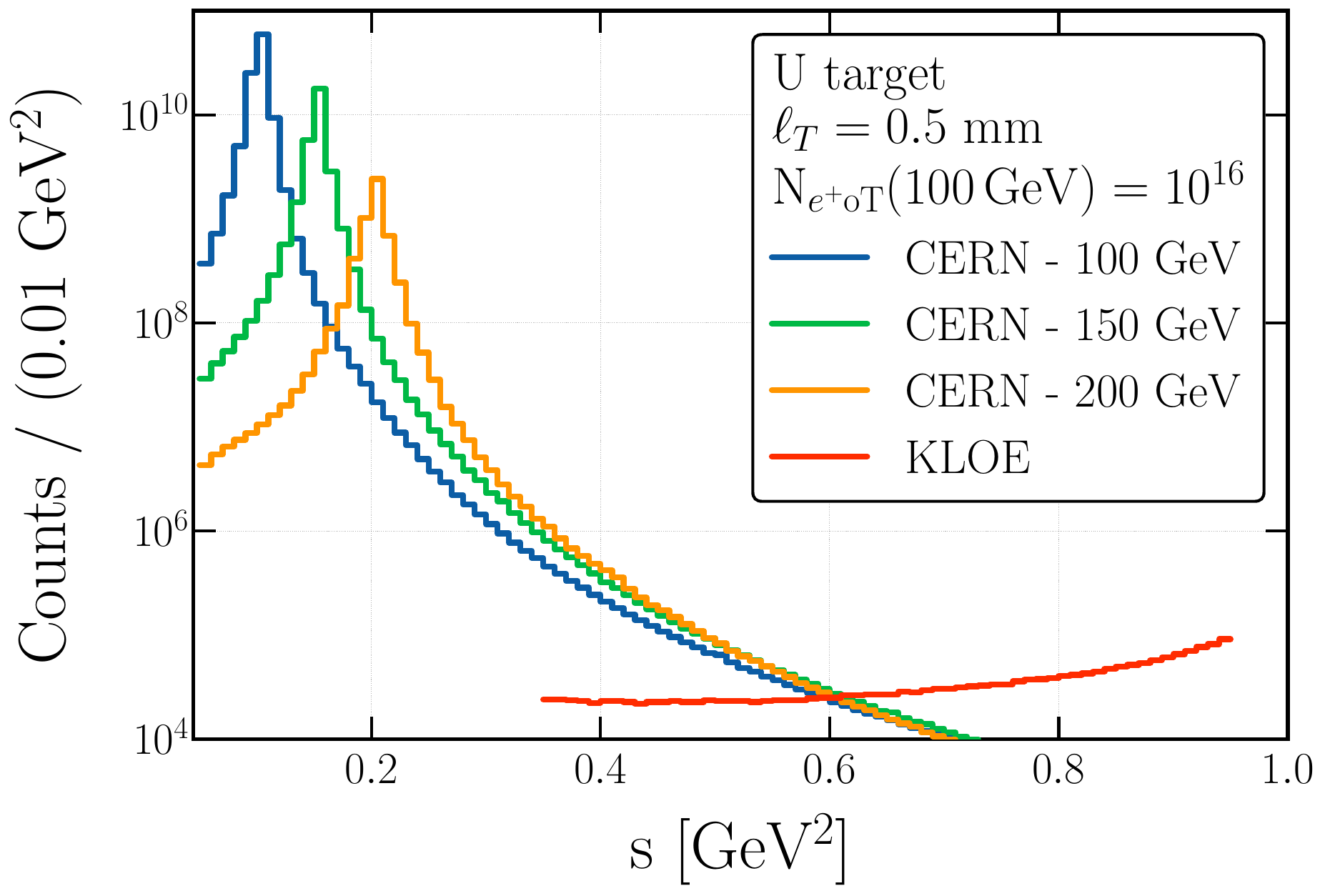}
    \caption{Number of $\mu^+\mu^-$ events produced with the CERN H4 line 
    assuming $N_{e^+\mathrm{oT}}\sim 10^{16}$ at $E_B=100\,$GeV (blue line)  and 
     an exponential  scaling
    of the beam intensity as given in \eq{eq:gatignon}
    for $E_B=150\,$GeV (green line) and $E_B=200\,$GeV (orange line).  
    The red line gives the number of 
    $\mu^+\mu^-(\gamma)$ detected by KLOE,  as reported 
    in Ref.~\cite{KLOE:2012anl}.}
    \label{fig:CERN}
\end{figure}

In the CERN North Experimental Area (NA),  positron beams of much higher energy can  be available, and it is then natural to ask how a one order of magnitude 
increase in energy would influence the measurement. 
The downside, however, is that the NA positron beams are tertiary beams. Spills of 
400\,GeV protons from the SPS first impinge on a beryllium target  producing all sorts of particles. Charged particles are deflected away, while secondary photons, and photons from $\pi_0$ decays,  pair produce $e^+e^-$  in a lead converter located downstream. 
Magnetic fields and collimators are then used for charge and momentum selection.  
As a result, a typical figure for the CERN  H4 line (serving the NA64 experiment)  
is $5 \cdot 10^6 \, e^+$oT/spill~\cite{l.gatignon}. Assuming 3500 spills/day this would corresponds to  $\sim 6.5\cdot 10^{12} \,$poT/yr, which is not sufficient to allow for a useful measurement. 
Nevertheless, it remains interesting to explore the energy dependence of the measurement. To this end, we assume an (unrealistic) $N_{e^+\mathrm{oT}}\sim 10^{16}$ for 
a beam energy $E_B=100\,$GeV, an energy spread $\sim 1\%$, and a $500\,\mu$m uranium target. 
To account for the scaling of the  beam intensity at  energies above 100\,GeV we adopt 
the Gatignon parametrization (see the technical note Ref.~\cite{Gatignon:2730780}),
which gives 
\begin{equation}
\label{eq:gatignon}
    \frac{N_{e^+\mathrm{oT}}(p')}
    {N_{e^+\mathrm{oT}}(p)} = e^{- \frac{B}{p_0}(p'-p)}\,,
\end{equation}
where $p_0$ is the energy of the primary protons ($400\,$GeV),
$B=10$, and $N_{e^+\mathrm{oT}}(100\, \mathrm{GeV})=10^{16}$.
Our results are depicted in Fig.~\ref{fig:CERN}
for three values of the positron energy $E_B=100,\,150,\,200\,$GeV
(respectively blue, green and orange lines).
The number of  $\mu^+\mu^-(\gamma)$ events detected 
by KLOE, as reported in Ref.~\cite{KLOE:2012anl}, 
is plotted for comparison (red line). 
In the range $4m_\pi^2 < s \lesssim 0.6\,\text{GeV}^2$
the statistical uncertainty on $\amhvp$ associated with our simulated dataset 
remains below $0.05\%$. Hence also in this case systematic uncertainties will 
play the dominant role.

It is intriguing to note that above $s \sim 0.5\, \text{GeV}^2$, increasing the beam energy does not result in a gain in statistics. Although
for a given value of $\sqrt{s}$ higher positron energies  allow probing smaller electron momenta, for which the momentum density 
distribution is less suppressed, the corresponding decrease in beam intensity, 
as parametrised in \eq{eq:gatignon}, more than offsets this advantage.
Thus, considering that  hadron contamination in the positron beams increases 
from  2--3\% at $E_B =100\,$GeV to more than 100\% at $E_B =200\,$GeV, see Ref.~\cite{Andreev:2023xmj}, lower beam energies turn out to be preferable. 
Finally, the result of a measurement with a more realistic 
$N_{e^+\mathrm{oT}}(100 \,\mathrm{GeV})=10^{13}$ can easily be inferred by rescaling 
the blue line in Fig.~\ref{fig:CERN} by a factor $10^{-3}$. 
In this case only the region $4 m_\pi^2 \lesssim s \lesssim 0.3\,\mathrm{GeV}^2 $
could be probed with sufficient statistical precision.  However, such a measurement 
would still be of remarkable interest, as it would be complementary to the 
ISR and energy scan techniques, that in this region exhibit reduced statistics. 

\Sec{Conclusions.}
The  novel strategy to measure $\shad$ that we have proposed could play a pivotal 
role in solving the conundrum of the  discordant determinations of 
the HVP contribution to the muon anomalous magnetic moment 
extracted from $e^+e^-\to\;$hadrons data. 
Presently, the most accurate estimates are obtained from 
measurements at $e^+e^- $ circular colliders. Two different techniques 
are employed to reconstruct the energy dependence of the cross-section: 
either the scanning method, in which the beam energies are varied, or 
the radiative return method, in which a hard photon is emitted from the initial 
state thus modifying the  c.m. energy of the collision.
We have shown that a different type of measurement can be carried out 
with positron beams in fixed target experiments.
A scan over the c.m. dependence of the cross section is automatically provided 
by leveraging the atomic electron velocities 
in high $Z$ target materials, like $^{238}$U. 
This method can provide statistically accurate measurements 
that are complementary to the ones at colliders, as the statistics will be particularly 
large in the low $\sqrt{s}$ region around the two pion threshold, where 
the ISR and scanning method are affected by the largest statistical uncertainties. 
We have studied the reach of the positron beam foreseen at JLab, and we have shown 
that with the anticipated beam parameters the whole region from the two-pion threshold 
up to above $\sqrt{s}\sim 1\,$GeV, that is the crucial one for accurate determinations 
of  $\amhvp$, can be fully covered.  Conversely, the 
positron beams available in the CERN NA do not have a sufficient intensity 
to cover in full the interesting range. 
Still, with realistic beam parameters $\shad(s)$ can be measured 
with good statistical accuracy in the low energy region 
$4 m_\pi^2 \lesssim s \lesssim 0.3\,\mathrm{GeV}^2$. 
The issue of experimental detection of $\mu^+\mu^-$ and $\pi\pi(\gamma)$ events with   
the related systematic uncertainties is beyond the scope of this work, 
and it has not been addressed in our study.
However, it is clear that to successfully carry out  
the measurement that we are proposing, it will be required, among other things, 
a good $\pi/\mu$ discrimination and an accurate reconstruction 
of the c.m. energy of the collision from the final states momenta.

\begin{acknowledgments}
\noindent
\textbf{Acknowledgments} --
We thank the authors of Ref.~\cite{Plestid:2024xzh} 
and in particular R. Plestid for contributing to identify 
an error in a first draft of this letter.
We warmly thank  Lau Gatignon and Johannes Bernhard for providing us with 
detailed information about the CERN NA beams,
and Eric Voutier for details on the CEBAF positron beam parameters. 
We acknowledge conversations with  M. Raggi and P. Valente. 
 E.N. acknowledges hospitality from the LAPth group in Annecy during the development of this work. 
F.A.A., G.G.d.C. and E.N. are supported in part by the INFN ``Iniziativa Specifica" Theoretical Astroparticle Physics (TAsP). F.A.A. received additional support from an INFN Cabibbo Fellowship, call 2022.
G.G.d.C. acknowledges LNF and Sapienza University for hospitality at various stages of this work. The work of E.N. is also supported  by the Estonian Research Council grant PRG1884.   Partial support from the CoE grant TK202 “Foundations of the Universe” and from the CERN and ESA Science Consortium of Estonia, grants RVTT3 and RVTT7, and  
 from the COST (European Cooperation in Science and Technology) Action COSMIC WISPers CA21106 are also acknowledged.
\end{acknowledgments}

\bibliography{biblio} % Produces the bibliography via BibTeX.

\end{document}